\documentclass[sigconf, nonacm]{acmart}

\AtBeginDocument{%
  \providecommand\BibTeX{{%
    \normalfont B\kern-0.5em{\scshape i\kern-0.25em b}\kern-0.8em\TeX}}}

\copyrightyear{2021}
\acmYear{2021}
\setcopyright{acmlicensed}
\acmConference[DeFi '21]{Proceedings of the 2021 ACM CCS Workshop on Decentralized Finance and Security}{November 19, 2021}{Virtual Event, Republic of Korea}
\acmBooktitle{Proceedings of the 2021 ACM CCS Workshop on Decentralized Finance and Security (DeFi '21), November 19, 2021, Virtual Event, Republic of Korea}
\acmPrice{15.00}
\acmDOI{10.1145/3464967.3488590}
\acmISBN{978-1-4503-8540-4/21/11}

\usepackage{array}
\usepackage{caption}
\usepackage{subcaption}

\usepackage{tikz}
\usetikzlibrary{positioning}

\theoremstyle{acmplain}

\begin{document}
\fancyhead{}

\title{Concentrated Liquidity in Automated Market Makers}

\author{Robin Fritsch}
\affiliation{%
  \institution{ETH Zürich}
}
\email{rfritsch@ethz.ch}

\begin{abstract}
We examine how the introduction of concentrated liquidity has changed the liquidity provision market in automated market makers such as Uniswap.
To this end, we compare average liquidity provider returns from trading fees before and after its introduction.
Furthermore, we quantify the performance of a number of fundamental concentrated liquidity strategies using historical trade data.
We estimate their possible returns and evaluate which perform best for certain trading pairs and market conditions.
\end{abstract}

\maketitle

\section{Introduction}
Recently, decentralized exchanges such as Uniswap, Balancer and Curve have seen tremendous growth in use. More than \$500 billion have been traded on all decentralized exchanges in total in the first half of 2021 \cite{theblock}.

Most decentralized exchanges are characterized by the following two innovative aspects.
Firstly, they allow users to trade in a fully decentralized manner and without having to give up custody of their tokens to a third party (as is the case with centralized exchanges).
Secondly, they use a completely new type of market design compared to traditional exchanges which operate central limit order books: an automated market maker.
Instead of matching orders of traders to liquidity providers' orders in the order book, they let traders swap tokens directly with a smart contract that holds the reserves of the liquidity providers.

When traders want to exchange a certain amount of one token, the amount of the second token they receive in return is determined by a publicly known trade function depending on the reserves in the contract.
In the case of constant product market makers like Uniswap, the amount is chosen such that the product of the reserves in the pool remains unchanged.
For every trade, traders pay a small trading fee which is distributed pro rata among all liquidity providers (LPs) contributing to the pool (i.e.\ LPs receive a fraction of the fees proportional to their share of the liquidity pool).

Recently, the release of Uniswap v3 has added a new feature to the market design: concentrated liquidity \cite{uniswapv3}.
While previously all liquidity providers provided liquidity indiscriminately (or \emph{passively}) across all prices, concentrated liquidity enables LPs to choose limited price ranges to provide liquidity to. This allows for active strategies similar to market making in traditional order book markets.
Better understanding this new market design is essential, as it seems to be emerging as the standard automated market maker. (Uniswap v3 became the largest decentralized exchange by trading volume in just the second month after its launch \cite{theblock}.)

In order to fully appreciate the benefits of concentrated liquidity, one needs to understand the inefficiency of passive liquidity in a constant product market maker like Uniswap.
Consider a pool of two stablecoins whose prices relative to each other always stay in the range $[0.99, 1.01]$. Assume a liquidity provider deposits 1000 units of each token when the ratio of the two prices is 1.
When the ratio of the prices is at the lower end of the range, the position will consist of about 995 and 1005 units of the two tokens, respectively, while it will be about 1005 and 995 units at the upper end. In particular, note that 995 units of each token will never be touched.
So the LP could have simply deposited 5 units of each token into the pool while pretending to have deposited 1000 units (and earning fees as if the latter was true).
This is the idea behind concentrated liquidity.
Note that since the LP only deposits a fraction of the tokens, the position can run out of one of the two tokens if the price leaves a certain price range. Then the position can no longer facilitate trading and will therefore no longer earn fees until the price returns to the price range.

In this paper, we study the impact the introduction of concentrated liquidity has had on the liquidity provision market.
We compare the returns of \emph{active LPs}, who choose certain price ranges in which they concentrate their liquidity, to the returns of \emph{passive LPs}, who provide liquidity across the complete price range.
We find that the introduction of concentrated liquidity has significantly reduced the earnings of passive LPs. While for the ETH-USDC pair active LPs now on average earn more in fees than passive LPs did before the introduction of concentrated liquidity, this is not the case for the stablecoin pair USDT-USDC.

Furthermore, we use historical trade data to quantify the performance of a number of fundamental active strategies utilizing concentrated liquidity.
We estimate their possible returns and evaluate which perform best for certain trading pairs and market conditions.
For a volatile pair like ETH-USDC, we find that active strategies can outperform passive liquidity, but it is not trivial to pick a good one as some active strategies underperform passive liquidity.
For the USDT-USDC stablecoin pair on the other hand, concentrated liquidity clearly outperforms passive liquidity and choosing the best strategy is simple.

\section{Related Work}
Automated market makers, i.e.\ mechanisms that automate the process of providing liquidity to a market, were first introduced with Hanson's logarithmic market scoring rule (LMSR) \cite{hanson03}.
Recently, a new type of automated market maker has been popularized by decentralized exchanges such as Uniswap: the constant function market maker (CFMM) \cite{angeris20oracles}.
General properties of this novel market design and how it behaves alongside traditional centralized exchanges, have been studied in \cite{angeris19} and \cite{aoyagi2021liquidity}.

The replication portfolio of liquidity provider positions in a constant product market maker is derived in \cite{clark20} and generalized to any CFMM in \cite{angeris2020curvature}. The latter work also analyses how the curvature of a CFMM, i.e.\ the choice the of the constant function, influences liquidity provider returns.
Finally, \cite{clark21} also finds the replicating portfolio for constant product market makers with concentrated liquidity.

A first attempt at quantifying the returns of liquidity providers from concentrated liquidity strategies was made in \cite{neuder21}.
However, the paper uses several strong simplifications, most notably the following two.
Firstly, it assumes that the strategies compete only with passive liquidity.
Secondly, the value of the liquidity position itself and thereby the loss a liquidity provider can suffer from changing prices is not taken into account.
Our analysis shows that both these aspects have a significant impact on the returns of liquidity providers and can therefore not be ignored.
More precisely, we find that evaluating active strategies against passive liquidity only massively overestimates returns, and that simply choosing a strategy that maximizes fees is not advisable, as the value of the liquidity position declines rapidly for such strategies.

\section{Concentrated Liquidity}

Technically, concentrated liquidity as introduced in \cite{uniswapv3} works as follows. Consider a pair of tokens $X$ and $Y$ where we measure the price of token $X$ in terms of token $Y$.
When providing liquidity for this pair, an LP chooses a price range $[p_a,p_b]$ in which they would like to provide liquidity. (We use the terms range and interval interchangeably in the following.) Together with the current price $p$, this determines the ratio of the two tokens the LP needs to deposit.
The exact amounts of both tokens the LP decides to deposit then determines the amount of liquidity $L$ provided to the interval.

When a trade occurs, it moves the current price of the pool.
All LPs earn a portion of the trading fees proportional to the amount of liquidity they provided to the interval by which the price moved.
Note that if the price moves outside the chosen range of an LP, the position consists of only one of the two tokens and stops earning fees.
Once the price returns to the range, the position earns fees again.

Providing liquidity to a price range can also be thought of as leveraging a passive LP position in the following sense: LPs only put up a fraction of the reserves which are used to calculate their proportion of the fees.
The smaller the chosen range, the higher the leverage and the larger the proportion of fees the LP can earn. However, smaller ranges also lead to greater losses for the LP when asset prices decline. We will demonstrate this with an example later.

\begin{figure}
    \centering
    \begin{tikzpicture}
      \draw[->] (0, 0) -- (5.5, 0) node[below] {$x$};
      \draw[->] (0, 0) -- (0, 5.5) node[left] {$y$};
      \draw[scale=1, domain=5/4:4, smooth, variable=\x, black] plot ({\x}, {5/\x});
      \node[fill, draw, circle, minimum width=3pt, inner sep=0pt] at (4, 5/4) {};
      \node[above right=2pt of {(4, 5/4)}] {$\left(x'(p_a), y'(p_a)\right)$};
      \node[fill, draw, circle, minimum width=3pt, inner sep=0pt] at (5/4, 4) {};
      \node[above right=2pt of {(5/4, 4)}] {$\left(x'(p_b), y'(p_b)\right)$};
      \draw[scale=1, domain=0:4-5/4, smooth, dashed, variable=\x, black] plot ({\x}, {5/(\x+5/4)-5/4});
      \node[fill, draw, circle, minimum width=3pt, inner sep=0pt] at (0, 4-5/4) {};
      \node[fill, draw, circle, minimum width=3pt, inner sep=0pt] at (4-5/4, 0) {};
    \end{tikzpicture}
    \caption{Real reserves (solid line) and virtual reserves (dashed line) of a concentrated liquidity position.}
    \label{fig:v3_liquidity}
\end{figure}
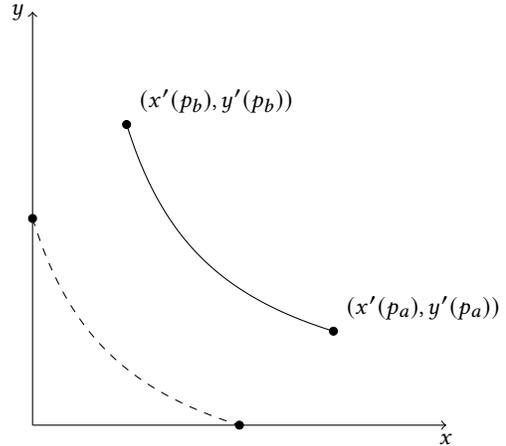

Consider a liquidity position in the range $[p_a,p_b]$ as shown in Figure \ref{fig:v3_liquidity}. Let $x'(p)$ and $y'(p)$ be the virtual reserves of the position at a price of $p\in [p_a,p_b]$ (i.e.\ the amounts of tokens used to calculate trades against the liquidity position). Then $x'(p) y'(p) = k = L^2$ and $y'(p)/x'(p) = p$. This implies $x'(p) = L/\sqrt{p}$ and $y'(p) = L\sqrt{p}$.
Hence, the real reserves $x(p)$ and $y(p)$ at price $p$ (i.e.\ the amounts of tokens the position actually consists of) can be calculated as follows.
\begin{align}
\begin{split}
    x(p) &= x'(p) - x'(p_b) = L\left(\frac{1}{\sqrt{p}} - \frac{1}{\sqrt{p_b}} \right) \\
    y(p) &= y'(p) - y'(p_a) = L\left(\sqrt{p} - \sqrt{p_a} \right)
\end{split}
\label{eq:concentrated_liquidity}
\end{align}

Note that the formulas above only hold for $p\in [p_a,p_b]$. 
When the price is not in this range, we have $x(p) = x(p_a)$ and $y(p) = 0$ for $p<p_a$ as well as $x(p) = 0$ and $y(p) = y(p_b)$ for $p>p_b$.

The equations \eqref{eq:concentrated_liquidity} can be used to calculate the current value of a liquidity position.
In our case, asset $Y$ will always be USDC, a stablecoin pegged to the US Dollar.
At price $p$, the liquidity position consists of $y(p)$ USDC and $x(p)$ units of the second token. The latter amount is worth $x(p)p$ in USDC, meaning the total value of the position is $y(p)+x(p)p$.

As an illustration, consider the following example.
Consider an ETH-USDC pool and assume the current price of ETH is 2000 USDC. Two liquidity providers $A$ and $B$ both start with a position worth 1000 USDC. LP $A$ chooses the range $[1818, 2200]$, while LP $B$ chooses $[1667, 2400]$.
According to \eqref{eq:concentrated_liquidity}, this means that they provide about 240 and 128 units of liquidity, respectively.
If ETH's price now falls to 1900, the position of LP $A$ will consist of 288.23 USDC and ETH worth 739.40 USDC which adds up to 967.63 USDC.
LP $B$'s position on the other hand consists of 354.54 USDC and ETH worth 617.27 USDC, i.e.\ 971.81 USDC in total. In particular, the LP with the smaller interval suffered a larger loss. On the other hand, LP A earns $240/128 = 1.875$ times more fees on trades occurring in the range $[1818, 2200]$ than LP B.

In Uniswap v3, LPs cannot choose liquidity range completely arbitrarily.
Instead, the price range is divided by so-called ticks which can be chosen as boundaries for liquidity ranges. The ticks are located at $1.0001^i$ for $i\in \mathbb{Z}$. For every pool, a tick spacing $t_s$ is specified, and only ticks with $t_s|i$ may be used.

\section{Strategies}

To get an idea of what the best strategies for active liquidity provision are and how much LPs can earn using concentrated liquidity strategies, we evaluate the following fundamental strategies.

\begin{description}
\item[No Liquidity Provision:]
Half of the portfolio is allocated to each of the two tokens. However, none of the liquidity is deposited into a pool.

\item[Passive Liquidity -- Uniswap v2:]
The liquidity is provided across the complete price range, just as in Uniswap v2.

\item[Fixed Interval -- $\mathbf{Fixed(a)}$:]
In the beginning, liquidity is provided to a symmetric interval around the current price, and this interval is never adjusted.
More specifically, the interval $(p(1+a)^{-1}, p(1+a))$ is chosen where $p$ is the current price and $a>0$ is a parameter that characterizes the strategy.
In other words, choosing $a=10\%$ means providing liquidity to the interval $\pm 10\%$ around the current price.
(By the choice of a symmetric interval, the LP is required to deposit equal amounts of both token.)
 
For ETH-USDC pools, we choose $a\in [0.6\%, 100\%]$ where $a$ grows in steps of $0.6\%$.
For USDT-USDC pools, we choose $a\in [0.1\%, 50\%]$ where $a$ grows in steps of $0.1\%$.

We choose these values to match the precision with which LPs can choose ranges in Uniswap v3. 
There, the tick spacing is 60 and 10 for ETH-USDC and USDT-USDC pools, respectively.
So for the latter pools, the smallest possible ratio between interval bounds is $1.0001^{10} \approx 1 + 0.1\%$.

\item[Resetting Interval -- $\mathbf{Reset(a,r)}$:]
In short, this strategy sets a symmetric liquidity range around the current price and adjusts it each time the price moves outside a certain range.
More precisely, the strategy depends on two parameters $a>0$ and $r>0$.
Initially, liquidity is provided to the interval $(p(1+a)^{-1}, p(1+a))$ around the current price $p$. Furthermore, a second interval $(p(1+r)^{-1}, p(1+r))$, the resetting interval, is chosen. As soon as the price leaves the resetting interval, the liquidity position is adjusted: Both the liquidity and the resetting interval are reset around the current price.
Since the LP may at this point own a larger value of one of the two tokens,
more liquidity may be provided on one side of the current price.
More precisely, the liquidity interval is split into the part above and the part below the current price, i.e.\ $(p(1+a)^{-1}, p)$ and $(p, p(1+a))$.
Providing liquidity to each of these intervals requires depositing exactly one of the two tokens. We demonstrate this strategy with an example below.

We choose the parameters $a$ and $r$ from the range $[1\%, 100\%]$ with steps of 1\% for ETH-USDC pools and from the range $[0.1\%, 5\%]$ with steps of 0.1\% for the USDT-USDC pools.

As an example, let $a=10\%$ and $r=5\%$. Consider an ETH-USDC pool and let the price initially be $p=2000$. Assuming the LP start with tokens worth 1000 USDC, 500 USDC worth of ETH and 500 USDC, this will result in about $L=240$ liquidity units in the range $[1818,2200]$. Now assume the price changes to $p=2100$. Then the liquidity position consists of 765.06 USDC and ETH worth about 252.87 USDC. The liquidity interval will now be reset to $[1909, 2310]$, leading to about 359 liquidity units being provided to the interval $[1909, 2100]$ as well as about 119 liquidity units to $[2100, 2310]$. 

\end{description}

\section{Evaluation Methodology}
To evaluate the strategies, we use real market data from a number of Uniswap pools. We consider pools for the pair ETH-USDC as well as for the stablecoin pair USDT-USDC, more precisely the following pools. (The percentage indicates the trading fee.)

\begin{itemize}\itemsep0em
    \item Uniswap v2 ETH-USDC 0.3\%
    \item Uniswap v2 USDT-USDC 0.3\%
    \item Uniswap v3 ETH-USDC 0.3\%
    \item Uniswap v3 ETH-USDC 0.05\%
    \item Uniswap v3 USDT-USDC 0.05\%
\end{itemize}

For the Uniswap v2 pools we use data from the first eight months of 2021, i.e.\ from January through August.
Since Uniswap v3 only launched on 5 May 2021, we use data from the months of June, July and August for these pools.

For all pools, we use hourly data. That means for every hour, we consider the trading volume over the past hour as well as the price and liquidity at the end of the hour. For v2 pools, the latter is simply the amount of liquidity in the pool. For the v3 pools, we use the liquidity in the tick of the price at the end of the hour.
In particular, this assumes that all volume over the past hour took place in the tick of the hourly closing price.

When simulating a certain strategy, we assume that the liquidity provided by the strategy is small relative to the total liquidity at that time.
Let $V_t$ be the trade volume over the past hour and $L_t$ be the actual total liquidity (in the tick of the closing price) at that time.
Then for a strategy providing $L$ liquidity units (in this tick), we simulate that it earns $V_t\cdot 0.3\% \cdot L/L_t$ (instead of $V_t\cdot 0.3\% \cdot L/(L_t+L)$) in fees if the pool charges $0.3\%$ in trading fees.

For each strategy, we assume that the liquidity position is worth 1 USDC in the beginning. We then track the following three metrics over the observation time period. 
\begin{description}
    \item[fees:] The non-compounding amount of fees earned, i.e. the sum of fees received over the time period. More precisely, this assumes that every time fees are received, they are converted to USDC and kept separately from the liquidity position.
    \item[value:] The liquidity position value without fees. This is the value of the liquidity position (initially worth 1 USDC) if the received fees are not added to the position (which is equivalent to the final value of the position if trading fees were zero).
    \item[total:] The value of the liquidity position including compounding fees at the end of the time period. Here, we assume that fees are added to the liquidity positions as soon as they are received.
\end{description}

\begin{figure*}[!ht]
\centering
\begin{subfigure}[b]{\textwidth}
    \centering
    \includegraphics[width=0.95\textwidth]{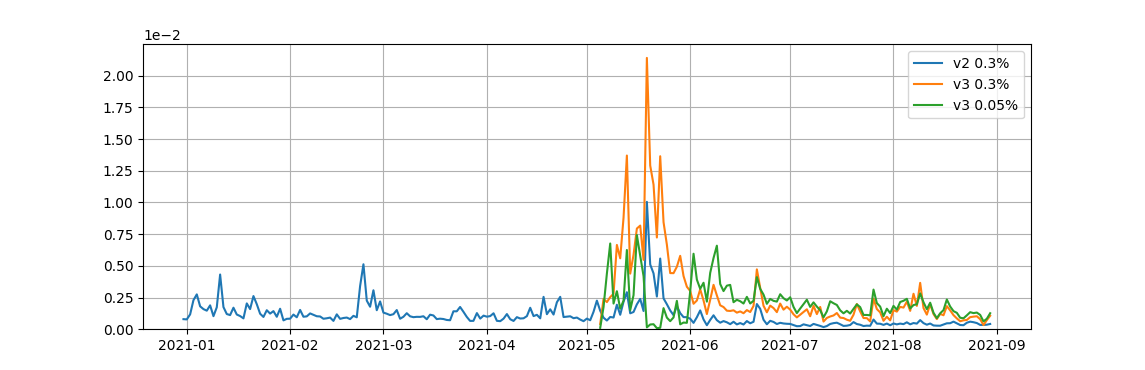}
    \caption{ETH-USDC pools}
    \label{fig:daily_returns_eth}
\end{subfigure}

\begin{subfigure}[b]{\textwidth}
    \centering
    \includegraphics[width=0.95\textwidth]{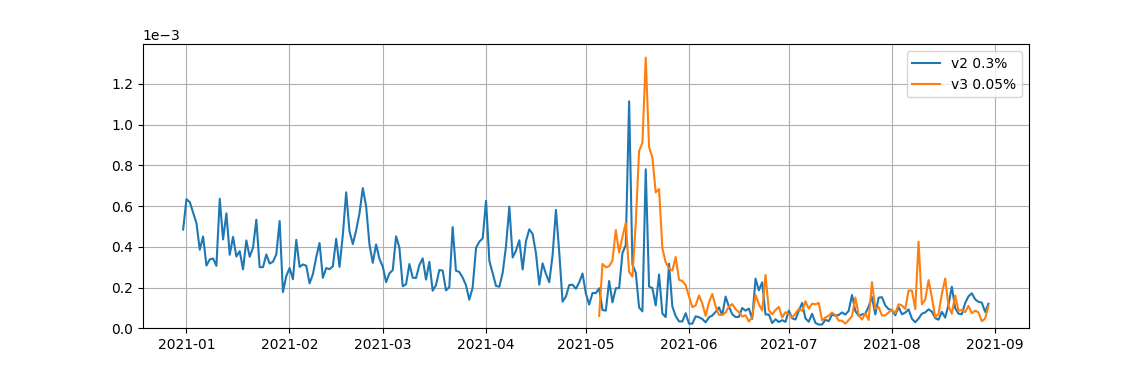}
    \caption{USDT-USDC pools}
    \label{fig:daily_returns_usdt}
\end{subfigure}
\caption{Daily liquidity provider returns from trading fees for two token pairs (in \$ per 1\$ of liquidity). Calculated as the daily amount of fees divided by the total value locked on that day.}
\label{fig:daily_lp_returns}
\end{figure*}

The most important metric clearly is total which shows how much a \$1 investment of the LP turns into over the observation period. We also track the two other metrics in order to get a picture of how much of the LP's overall profit or loss stems from earned fees vs.\ the changing value of the position.

Note that fees in most AMMs including Uniswap v2 are automatically compounding, while this is not the case for Uniswap v3.

\section{Results}

Before looking at the strategies in more detail, we first compare the overall returns from trading fees of liquidity providers before and after the introduction of concentrated liquidity.
Figures \ref{fig:daily_returns_eth} and \ref{fig:daily_returns_usdt} show the daily liquidity provider returns in ETH-USDC and USDT-USDC pools, respectively.

These are calculated by simply dividing the total amount of fees over a day by the total value locked in the pool. For Uniswap v3, this results in the average liquidity provider return, as all LPs earn differently according to their chosen price ranges.
Note that (only in this first step) we only take the daily fees earned into account and not the daily change in value of the liquidity position.


Both plots show high spikes in the month of June, a period that was characterized by high volatility and trading volume. (The price of ETH fell more than 50\% in two weeks during that time).
To examine the effects of concentrated liquidity, we ignore this period and compare the time before Uniswap v3 (01.01.21 -- 30.04.21) to the time after v3's launch (01.06.21 -- 31.08.21). The average daily liquidity provider returns over these two time periods are shown in Table \ref{tab:avg_LP_returns}.

\begin{table}[!h]
\setlength\tabcolsep{3pt}
\begin{center}
\begin{tabular}{ |c|c|c| }
 \hline
 Pool & 01.01.21 -- 30.04.21 & 01.06.21 -- 31.08.21 \\
 \hline
  ETH-USDC v2 0.3\% & $1.30\times 10^{-3}$ & $0.50\times 10^{-3}$ \\
 \hline
  ETH-USDC v3 0.3\% & - & $1.51\times 10^{-3}$ \\
 \hline
  ETH-USDC v3 0.05\% & - & $2.12\times 10^{-3}$ \\
 \hline
  USDT-USDC v2 0.3\% & $3.52\times 10^{-4}$ & $0.83 \times 10^{-4}$\\
 \hline
  USDT-USDC v3 0.05\% & - & $1.02\times 10^{-4}$\\
 \hline
\end{tabular}
\end{center}
\caption{Average daily liquidity provider returns from trading fees before and after the introduction of concentrated liquidity (in \$ per 1\$ of liquidity).}
\label{tab:avg_LP_returns}
\end{table}

\begin{table*}[!ht]
\setlength\tabcolsep{3pt}
\centering
\begin{subtable}{\textwidth}
\centering
\begin{tabular}{ |c|c|cc>{\bfseries}c|c|cc>{\bfseries}c|c|cc>{\bfseries}c| }
 \hline
  & \multicolumn{4}{|c|}{01.06.21 -- 30.06.21} & \multicolumn{4}{|c|}{01.07.21 -- 31.07.21} & \multicolumn{4}{|c|}{01.08.21 -- 31.08.21}\\
 Strategy & Parameters & fees & value & total & Parameters & fees & value & total & Parameters & fees & value & total \\
 \hline
  No liquidity provision & - & 0.000 & 0.927 & 0.927 & - & 0.000 & 1.061 & 1.061 & - & 0.000 & 1.173 & 1.173 \\
 \hline
  Passive liquidity & - & 0.010 & 0.924 & 0.934 & - & 0.006 & 1.060& 1.067 & - & 0.007 & 1.160 & 1.167\\
 \hline
  Best Fixed (total) & 48.0\% & 0.053 & 0.911 & 0.967 & 3.0\% & 0.114 & 1.007 & 1.130 & 99.6\% & 0.022 & 1.129 & 1.153 \\
 \hline
  Worst Fixed (total) & 18.0\% & 0.056 & 0.890 & 0.944 & 99.6\% & 0.021 & 1.055 & 1.079 & 0.6\% & 0.033 & 1.001 & 1.035 \\
 \hline
  Best Fixed (fees) & 0.6\% & 0.112 & 0.855 & 0.956 & 3.0\% & 0.114 & 1.007 & 1.130 & 3.0\% & 0.055 & 1.007 & 1.064\\
  \hline
  Best Reset (total) & 1.2\%, 15.0\% & 0.300 & 0.738 & 1.059 & 3.0\%, 30.6\% & 0.114 & 1.007 & 1.130 & 0.6\%, 5.4\% & 0.416 & 0.894 & 1.383\\
 \hline
  Worst Reset (total) & 0.6\%, 0.6\% & 0.370 & 0.114 & 0.307 & 0.6\%, 0.6\% & 0.403 & 0.258 & 0.580 & 0.6\%, 0.6\% & 0.641 & 0.346 & 0.948\\
 \hline
   Best Reset (fees) & 1.2\%, 1.2\% & 0.492 & 0.209 & 0.576 & 0.6\%, 1.2\% & 0.489 & 0.316 & 0.750 & 0.6\%, 1.8\% & 0.684 & 0.489 & 1.223\\
 \hline
\end{tabular}
\caption{Uniswap v3 ETH-USDC 0.3\% pool data.}
\label{tab:v3_ETH_USDC_30}
\end{subtable}
\newline
\vspace*{0.3cm}
\newline
\begin{subtable}{\textwidth}
\centering
\begin{tabular}{ |c|c|cc>{\bfseries}c|c|cc>{\bfseries}c|c|cc>{\bfseries}c| }
 \hline
  & \multicolumn{4}{|c|}{01.06.21 -- 30.06.21} & \multicolumn{4}{|c|}{01.07.21 -- 31.07.21} & \multicolumn{4}{|c|}{01.08.21 -- 31.08.21} \\
 Strategy & Parameters & fees & value & total & Parameters & fees & value & total & Parameters & fees & value & total \\
 \hline
  No liquidity provision & - & 0.000 & 0.927 & 0.927 & - & 0.00 & 1.062 & 1.062 & - & 0.000 & 1.173 & 1.173\\
 \hline
  Passive liquidity & - & 0.020 & 0.924 & 0.943 & - & 0.011 & 1.060& 1.072 & - & 0.015 & 1.160 & 1.176\\
 \hline
  Best Fixed (total) & 1.2\% & 0.296 & 0.856 & 1.150 & 4.2\% & 0.189 & 1.010 & 1.221 & 35.4\% & 0.108 & 1.082 & 1.198\\
 \hline
  Best Reset (total) & 1.2\%, 3.0\% & 1.178 & 0.321 & 1.760 & 0.6\%, 1.2\% & 0.946 & 0.301 & 1.889 & 0.6\%,0.6\% & 1.583 & 0.328 & 5.663 \\
 \hline
\end{tabular}
\caption{Uniswap v2 ETH-USDC 0.3\% pool data.}
\label{tab:v2_ETH_USDC_30}
\end{subtable}
\caption{Performances of strategies for the ETH-USDC pair. (All performance metrics are rounded to three digits after the comma.)}
\label{tab:ETH_USDC_results}
\end{table*}

\begin{table*}[!ht]
\setlength\tabcolsep{3pt}
\centering
\begin{subtable}{\textwidth}
\centering
\begin{tabular}{ |c|c|cc>{\bfseries}c|c|cc>{\bfseries}c|c|cc>{\bfseries}c| }
 \hline
  & \multicolumn{4}{|c|}{01.06.21 -- 30.06.21} & \multicolumn{4}{|c|}{01.07.21 -- 31.07.21} & \multicolumn{4}{|c|}{01.08.21 -- 31.08.21} \\
 Strategy & Parameters & fees & value & total & Parameters & fees & value & total & Parameters & fees & value & total \\
 \hline
  No liquidity provision & - & 0.0000 & 0.9999 & 0.9999 & - & 0.0000 & 1.0000 & 1.0000 & - & 0.0000 & 1.0001 & 1.0001\\
 \hline
  Passive liquidity & - & 0.0000 & 0.9999 & 0.9999 & - & 0.0000 & 1.0000 & 1.0000 & - & 0.0000 & 1.0001 & 1.0001 \\
 \hline
  Best Fixed (total) & 0.1\% & 0.0034 & 0.9999 & 1.0033 & 0.1\% & 0.0027 & 1.0000 & 1.0027 & 0.1\% & 0.0039 & 1.0001 & 1.0040\\
 \hline
  Best Reset (total) & 0.1\%, 0.1\% & 0.0034 & 0.9999 & 1.0033 & 0.1\%, 0.1\% & 0.0027 & 1.0000 & 1.0027 & 0.1\%,0.1\% & 0.0039 & 1.0001 & 1.0040\\
 \hline
\end{tabular}
\caption{Uniswap v3 USDT-USDC 0.05\% pool data.}
\label{tab:v3_USDT_USDC_05}
\end{subtable}
\newline
\vspace*{0.3cm}
\newline
\begin{subtable}{\textwidth}
\centering
\begin{tabular}{ |c|c|cc>{\bfseries}c|c|cc>{\bfseries}c|c|cc>{\bfseries}c| }
 \hline
  & \multicolumn{4}{|c|}{01.06.21 -- 30.06.21} & \multicolumn{4}{|c|}{01.07.21 -- 31.07.21} & \multicolumn{4}{|c|}{01.08.21 -- 31.08.21} \\
 Strategy & Parameters & fees & value & total & Parameters & fees & value & total  & Parameters & fees & value & total \\
 \hline
  No liquidity provision & - & 0.0000 & 1.0003 & 1.0003 & - & 0.0000 & 0.9984 & 0.9984 & - & 0.0000 & 1.0002 & 1.0002 \\
 \hline
  Passive liquidity & - & 0.0024 & 1.0003 & 1.0027 & - & 0.0024 & 0.9984 & 1.0009 & - & 0.0030 & 1.0002 & 1.0032 \\
 \hline
  Best Fixed (total) & 0.1\% & 1.7937 & 1.0002 & 5.8659 & 0.1\% & 1.250 & 0.9971 & 3.4381 & 0.1\% & 1.9735 & 1.0002 & 7.0900 \\
 \hline
  Best Reset (total) & 0.2\%, 0.2\% & 1.8743 & 0.9233 & 6.2943 & 0.2\%, 0.4\% & 1.7964 & 0.9405 & 5.9161 & 0.1\%,0.2\% & 2.6301 & 0.8421 & 14.7249\\
 \hline
\end{tabular}
\caption{Uniswap v2 USDT-USDC 0.3\% pool data.}
\label{tab:v2_USDT_USDC_30}
\end{subtable}
\caption{Performances of strategies for the USDT-USDC pair. (All performance metrics are rounded to four digits after the comma.)}
\label{tab:USDT_USDC_results}
\end{table*}

We see that the returns of passive liquidity providers (in Uniswap v2 pools) have decreased significantly with the introduction of contracted liquidity.
A second key takeaway is that active LPs (i.e.\ LPs in Uniswap v3 pools) are on average earning significantly more fees than passive LPs in the v2 pool for the ETH-USDC pair.
For the USDT pools on the other hand, the average v3 LP does not earn a lot more fees than a v2 LP. We will later discuss that the latter fact is most likely caused by suboptimal trading.
Finally, we note that in the stablecoin pools both passive and active LPs currently earn significantly less than LPs did prior to the introduction of concentrated liquidity.
For the ETH-USDC pair on the other hand, active LPs now earn more in fees than passive LPs did before concentrated liquidity.


Now we turn to the performances of the examined strategies. These can be found in Tables \ref{tab:ETH_USDC_results} and \ref{tab:USDT_USDC_results}.
We evaluate all strategies against trade data from pools with and without concentrated liquidity (i.e.\ Uniswap v2 and v3 pools).
For the active strategies, we show the parameters that performed best and worst in terms of the total metric, as well as the parameters that earned the most fees.

For the ETH-USDC pools, we see in Table \ref{tab:v3_ETH_USDC_30} that the best active strategies can significantly outperform passive liquidity. Of course, this may come with higher risk and the optimal parameters have been chosen in hindsight. Furthermore, there were also active strategies that underperformed passive liquidity in every time period.

We now discuss the individual strategies in more detail.
For fixed interval strategies, we see (specially by comparing July and August) that the optimal strategy depends on how much the price of ETH fluctuates:
If the price of ETH did not change a lot during the time period (as in July), the best strategy was to choose a small liquidity range and the worst was to choose a large one. When ETH's price changed a lot (as in August), it was the other way around.

For resetting interval strategies however, we can see a common pattern even across periods with different price trends.
During our (relatively short) observation period, the best strategy tended to be to set a small liquidity interval and a large resetting interval.
It is also notable that the worst resetting strategy was always to choose the smallest liquidity and resetting interval from our parameter set.

We also see in the last row of Table \ref{tab:v3_ETH_USDC_30} that the amount of fees earned is unsurprisingly maximized by choosing a small range and frequently resetting it as the prices changes. Then it is possible to earn close to 50\% in fees in a month. This however, does not lead to a high overall return, as the value of the liquidity position itself declines drastically. So picking a strategy that maximizes the amount of fees earned is not advisable.

Furthermore, comparing Table \ref{tab:v3_ETH_USDC_30} to Table \ref{tab:v2_ETH_USDC_30} shows that evaluating active strategies assuming that all other liquidity is passive, as done in previous work \cite{neuder21}, massively overestimates the returns.
This can be seen even more clearly for the USDT-USDC pools when comparing Tables \ref{tab:v3_USDT_USDC_05} and \ref{tab:v2_USDT_USDC_30}.

For the USDT-USDC pools, Table \ref{tab:v3_USDT_USDC_05} shows that the optimal strategy is unsurprisingly to provide liquidity in the smallest possible interval around 1.
An active LP then earned about 0.3\% in fees each month.
It is notable that passive liquidity earns almost zero fees when competing with concentrated liquidity in the v3 pool. However, at the same time passive LPs in the v2 pool still earned 0.24\% in fees each month. 
The difference can not be explained by the fact that the v2 pool charges 0.3\% in fees while the v3 pool only charges 0.05\%. In fact, passive LPs in the v3 pool earn less than $2\times 10^{-6}$ in fees in each month.

One would expect that passive liquidity earns equal returns in the v2 and v3 pools.
Indeed, for any two pools with the same current price, any optimally executed trade leads to the same amount of fees (relative to the size of the position) for passive liquidity in both pools.
It seems unlikely that the extra fees in the v2 pool stem solely from arbitragers balancing the pools.
This suggests that a significant amount of trades are being executed suboptimally using the v2 pool instead of the v3 pool. We expect this to diminish over time leading to almost zero earnings for passive LPs in all stablecoin pools.

\section{Conclusion}
For stablecoin pairs, our analysis shows that LPs can achieve significantly higher returns using concentrated liquidity strategies.
This suggests that concentrated liquidity will in the long term out compete passive liquidity for these pairs.
Currently, passive liquidity seems to being subsidized by suboptimal trading which should diminish over time.

For volatile pairs such as ETH-USDC, active LPs do on average earn more fees.
However, picking a strategy that outperforms passive liquidity is non-trivial.

In general, the introduction of concentrated liquidity has lead to a significant decrease in fee revenue for passive LPs.
While for the ETH-USDC pair, active LPs now earn more fees (accompanied by a higher risk) than passive LPs did in pre Uniswap v3 times, active LPs actually earns less from fees now than passive LPs used to for the USDT-USDC pair.
In the end, the main beneficiary of the introduction of concentrated liquidity may be traders who profit from reduced slippage.

It will be interesting to see how the liquidity provision market evolves over time and to broaden the analysis in the future when data covering a longer period of time becomes available.


\bibliographystyle{ACM-Reference-Format}
\bibliography{references}

\end{document}